\documentstyle{mn}
\newif\ifAMStwofonts
\RequirePackage{graphicx}
\def\hr{\hbox{$^{\rm h}$}}                 
\def\mm{\hbox{$^{\rm m}$}}                 
\def\fday{\hbox{$.\!\!^{\rm d}$}}          
\ifoldfss
  \ifCUPmtlplainloaded \else
    \NewTextAlphabet{textbfit} {cmbxti10} {}
    \NewTextAlphabet{textbfss} {cmssbx10} {}
    \NewMathAlphabet{mathbfit} {cmbxti10} {} 
    \NewMathAlphabet{mathbfss} {cmssbx10} {} 
  \fi
  \ifAMStwofonts
    \ifCUPmtlplainloaded \else
      \NewSymbolFont{upmath} {eurm10}
      \NewSymbolFont{AMSa} {msam10}
      \NewMathSymbol{\upi}     {0}{upmath}{19}
      \NewMathSymbol{\umu}     {0}{upmath}{16}
      \NewMathSymbol{\upartial}{0}{upmath}{40}
      \NewMathSymbol{\leqslant}{3}{AMSa}{36}
      \NewMathSymbol{\geqslant}{3}{AMSa}{3E}

       \let\ge=\geqslant
    \fi
  \fi
\fi 

\ifnfssone
  \newmathalphabet{\mathit}
  \addtoversion{normal}{\mathit}{cmr}{m}{it}
  \addtoversion{bold}{\mathit}{cmr}{bx}{it}
  \newmathalphabet{\mathbfit} 
  \addtoversion{normal}{\mathbfit}{cmr}{bx}{it}
  \addtoversion{bold}{\mathbfit}{cmr}{bx}{it}
  \newmathalphabet{\mathbfss} 
  \addtoversion{normal}{\mathbfss}{cmss}{bx}{n}
  \addtoversion{bold}{\mathbfss}{cmss}{bx}{n}
  \ifAMStwofonts
    \ifCUPmtlplainloaded \else
      \UseAMStwoboldmath
      \makeatletter
      \new@mathgroup\upmath@group
      \define@mathgroup\mv@normal\upmath@group{eur}{m}{n}
      \define@mathgroup\mv@bold\upmath@group{eur}{b}{n}
      \edef\UPM{\hexnumber\upmath@group}
      \new@mathgroup\amsa@group
      \define@mathgroup\mv@normal\amsa@group{msa}{m}{n}
      \define@mathgroup\mv@bold\amsa@group{msa}{m}{n}
      \edef\AMSa{\hexnumber\amsa@group}
      \makeatother
      \mathchardef\upi="0\UPM19
      \mathchardef\umu="0\UPM16
      \mathchardef\upartial="0\UPM40
      \mathchardef\leqslant="3\AMSa36
      \mathchardef\geqslant="3\AMSa3E

       \let\ge=\geqslant
    \fi
  \fi
\fi 

\ifnfsstwo
  \DeclareMathAlphabet{\mathbfit}{OT1}{cmr}{bx}{it}
  \SetMathAlphabet\mathbfit{bold}{OT1}{cmr}{bx}{it}
  \DeclareMathAlphabet{\mathbfss}{OT1}{cmss}{bx}{n}
  \SetMathAlphabet\mathbfss{bold}{OT1}{cmss}{bx}{n}
  \ifAMStwofonts
    \ifCUPmtlplainloaded \else
      \DeclareSymbolFont{UPM}{U}{eur}{m}{n}
      \SetSymbolFont{UPM}{bold}{U}{eur}{b}{n}
      \DeclareSymbolFont{AMSa}{U}{msa}{m}{n}
      \DeclareMathSymbol{\upi}{0}{UPM}{"19}
      \DeclareMathSymbol{\umu}{0}{UPM}{"16}
      \DeclareMathSymbol{\upartial}{0}{UPM}{"40}
      \DeclareMathSymbol{\leqslant}{3}{AMSa}{"36}
      \DeclareMathSymbol{\geqslant}{3}{AMSa}{"3E}

       \let\ge=\geqslant
    \fi
  \fi
\fi 

\ifCUPmtlplainloaded \else
  \ifAMStwofonts \else 
    \def\upi{\pi}
    \def\umu{\mu}
    \def\upartial{\partial}
  \fi
\fi

\title{TYC 1031 01262 1: The First Known Galactic Eclipsing Binary with a
Type II Cepheid Component}
\author[Antipin, Sokolovsky, Ignatieva]
{
S.V. Antipin$^{1,2}$\thanks{E-mail: antipin@sai.msu.ru (SVA);
idkfa@sai.msu.ru (KVS)},
K.V. Sokolovsky$^{1,3}$\footnotemark[1],
T.I. Ignatieva$^1$\\
$^1$Sternberg Astronomical Institute, 13, University Ave.,
Moscow 119992, Russia\\
$^2$Institute of Astronomy, Russian
Academy of Sciences, 48, Pyatnitskaya Str., Moscow 119017, Russia\\
$^3$Astro Space Center, Lebedev Physical Institute, Russian
Academy of Sciences, Profsoyuznaya str., 84/32, Moscow 117997,
Russia
}

\date{Accepted 2007 May 3.
      Received 2007 May 3;
      in original form 2007 April 28}

\pagerange{\pageref{firstpage}--\pageref{lastpage}}
\pubyear{0000}

\begin{document}

\maketitle

\label{firstpage}

\begin{abstract}
We present the discovery and CCD observations of the first
eclipsing binary with a Type~II Cepheid component in our Galaxy.
The pulsation and orbital periods are found to be 4.1523 and 51.38
days, respectively, i.e. this variable is a system with the
shortest orbital period among known Cepheid binaries. Pulsations
dominate the brightness variations. The eclipses are assumed to
be partial. The EB-subtype eclipsing light curve permits to
believe that the binary's components are non-spherical.
\end{abstract}

\begin{keywords}
variables: Cepheids -- binaries: eclipsing -- binaries: close
\end{keywords}

\section{Introduction}

At present, we know three Type II Cepheids which are components of
short-period binary systems. They are TX Del (the orbital period
of 133 days),  IX Cas (110 days) (Harris \& Welch, 1989), and AU
Peg (53.3 days) (Vink\'{o} et al., 1993). The binary separations
in these systems are near the possible minimum for the size of a
Type II Cepheid. Tidal forces influence the pulsating components
of such systems, and it is even possible that their pulsations are
excited by tidal effects (Harris et al., 1984). The orbital
inclination of these binary systems relative to the terrestrial
observer does not make them eclipsing.

Three long period eclipsing binary systems with Type I or II
Cepheid components have been discovered in the Large Magellanic
Cloud by OGLE and MACHO microlensing photometry programs (Udalski
et al., 1999; Alcock et al., 2002; Lepischak et al., 2004). The
importance of search for Cepheids -- members of eclipsing
binaries is considerable. If an eclipsing binary is a double-line
system, then mass, radius, and luminosity of the components can be
directly determined from the analysis of the light and radial
velocity curves. This gives us an opportunity to investigate the
structure and evolution of Cepheids, to test the theories of
pulsation and to calibrate independently the scale of distances
and the Hubble constant (Guinan et al., 2005; Guinan \& Engle,
2006).

In the past, BM Cas was considered a candidate  eclipsing binary
with a Cepheid component in the Galaxy, but detailed
investigation by Fernie \& Evans (1997) revealed that
out-of-eclipse light variations were non-periodic, inconsistent
with the Cepheid classification.

In this paper, we announce the discovery and present our CCD
photometry of the first eclipsing Type II Cepheid in the Galaxy
-- a component of the binary with the shortest orbital period
among similar systems.

\section{Observations and Analysis}

The variability of TYC~1031~01262~1 = ASAS 182611+1212.6
($\alpha$ = 18\hr26\mm11\fs50, $\delta$ =
+12\degr12\arcmin34\farcs8, J2000.0) was discovered several years
ago in the ASAS-3 survey (Pojma\'{n}ski et al., 2005). The
variability was independently discovered by the authors on Moscow
archive plates taken with the 40-cm astrograph in Crimea. The
photographic phased light curve based on 120 eye estimates is
given in Fig.~1. Moreover, observations of TYC~1031~01262~1 are
contained in the NSVS database (Wo\'{z}niak et al., 2004). In all
the three cases, the data permit us to consider the new variable
as a Cepheid with some peculiarity that is rather similar to
multiperiodicity of unknown nature. However, the available data
did not allow us to completely explain the observed light
variations. For this reason, we started observations of the star
in 2004 with the 50-cm Maksutov telescope of the Crimean
Laboratory (Sternberg Astronomical Institute) equipped with a
Pictor~416XTE CCD camera and Johnson $V$ filter.

\begin{figure}
 \centerline{\includegraphics[angle=0,width=83mm]{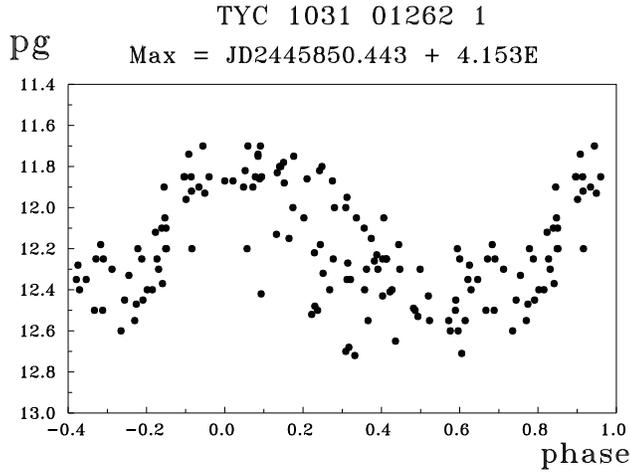}}
 \caption{The photographic phased light curve.}
\end{figure}

The observations continued for three years: 729 images on 17
nights in 2004 (July 4--29, JD2453191--216), 560 images on 13
nights in 2005 (July 1--19, JD2453553--571), and additional 367
images on 24 nights in 2006 (July 4 -- August 2, JD2453921--950)
were obtained. The images were dark subtracted, flat-fielded and
analyzed with the aperture photometry software $Winfits$
developed by V.P. Goranskij. The comparison stars are marked in
Fig.\,2. To improve the accuracy of our photometry and to
evaluate uncertainties, we used two comparison stars and averaged
the obtained differential magnitudes having in mind that
$V_{comp1}-V_{comp2}=0\fm123$. The accuracy of our photometry is
about 0\fm017. The data is available upon request. Combination of
our results with the ASAS-3 data permits us to propose a suitable
explanation of the variations observed for TYC~1031~01262~1. Note
that we (1) used the ASAS-3 observations taken from the official
web site of the project {\it before} the 2006 RAID problem,
(2)~made use of the third of the five columns of ASAS-3 data
which corresponds to the aperture 4 pixels in diameter, and
(3)~assumed $V_{comp1}=13\fm12$ to make our $V$-band observations
agree with the ASAS-3 ones.

\begin{figure}
 \centerline{\includegraphics[angle=0,width=83mm]{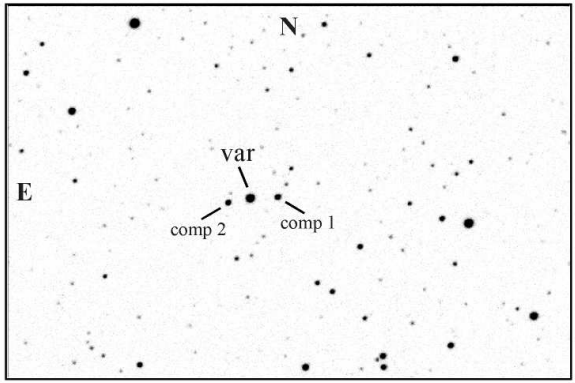}}
 \caption{The $V$-band image ($12' \times 8'$) of
the field of TYC~1031~01262~1. The comparison stars are marked.}
\end{figure}

Figure 3 shows the suggested solution of the observed variability
of TYC~1031~01262~1. Cepheid oscillations with the elements
$$Max = HJD 2453196.529 + 4\fday 1523 \times E $$
dominate in the light curve. Then the data were whitened for
pulsations and we found that the residuals could be well
described by an eclipsing (EB-subtype) curve with the following
light elements:
$$Min = HJD 2453571.36 + 51\fday 38 \times E. $$
The photographic and NSVS observations do not contradict the
proposed interpretation. However, in these cases, the eclipsing
light curves  after whitening the data for pulsations are rather
noisy.

\begin{figure}
 \centerline{\includegraphics[angle=0,width=83mm]{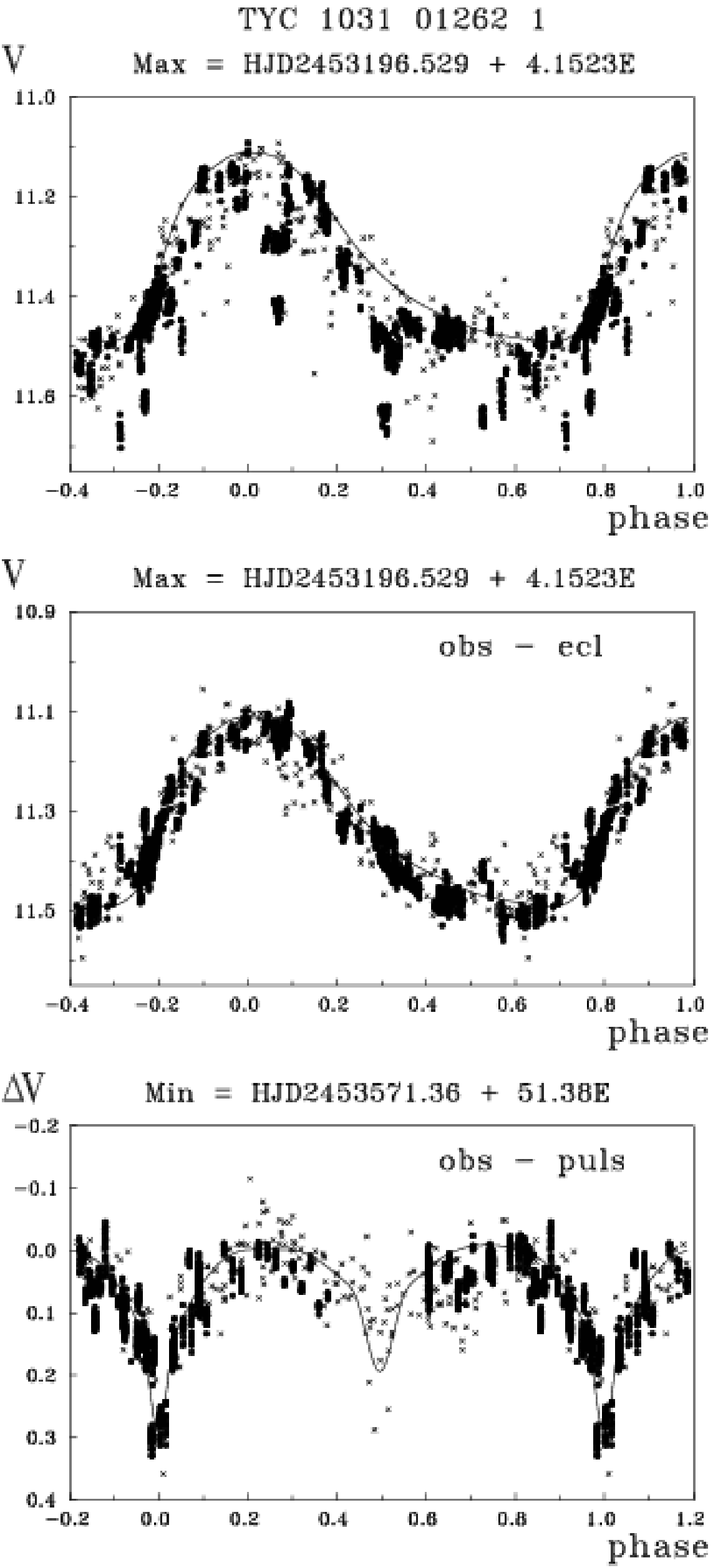}}
 \caption{The phased light curves in the $V$ band.
The ASAS-3 data are shown as crosses; the filled circles are our
CCD observations. The averaged phased light curves that have been
used for whitening to make the other periodicity evident are
plotted as solid curves.}
\end{figure}

\section{Discussion and Conclusions}

TYC 1031 01262 1 is definitely a Type II Cepheid. Using the
$K$-band period-luminosity relations for Galactic classical
Cepheids by Berdnikov et al. (1996):
$$M_K = -5.462-3.517 \times (lgP-1),$$
$K_S=9.525$ (from the 2MASS catalogue) and the galactic latitude
$b=11\degr$, considering that extinction in $K$ band is not
significant (less than 0.05 mag for the variable), we can estimate
the distance from the galactic plane as $Z \ge 1$ kpc, in
contradiction with the DCEP classification. Note that even in the
case of extinction being actually large the last conclusion will
not change.

For now, we have the following multi-colour photometry of the
system: $B_T=12.034\pm0.124$ and $V_T=11.431\pm0.109$ from Tycho-2
catalogue (H\o g et al., 2000); average $V=11.353\pm0.016$ and
$I=10.530\pm0.012$ from TASS observations (Droege et al., 2006);
$\langle V \rangle =11.376\pm0.010$ from ASAS-3 data;
$J=9.998\pm0.020$, $H=9.630\pm0.026$ and $K_S=9.525\pm0.020$ taken
on JD 2451613.0243 (from 2MASS, Cutri et al., 2003). The colours
do not differ noticeably from those of a single Type II Cepheid.
Direct spectroscopy is required to determine the spectral types of
the components.

Our detection of eclipses and observations solely in the $V$
filter are only the first step in the investigation of this unique
system. It is necessary to continue observations of the variable
star, carry out multi-colour photometry and spectroscopy to
confirm the derived orbital period in radial-velocity variations
and to determine parameters of the binary system. The most
similar, non-eclipsing binary system with a Cepheid component,
AU~Peg, shows a considerably larger amplitude of the orbital
radial velocity variation than that for the pulsations, $K=44.7$
km/s versus the pulsational semi-amplitude of about 13~km/s
(Samus et al., 1997). We can expect a similar situation in the
case of TYC~1031~01262~1.

Judging from their small amplitude, the eclipses are partial, and
therefore, the eclipse depth must depend on the Cepheid's radius,
i.e. on the phase of the pulsation period. This effect is
slightly noticeable in our observations, but it must be confirmed
and studied in detail by subsequent photometry. Moreover, we
would like to turn the reader's attention to the fact that the
eclipsing light curve being of the $\beta$~Lyrae type gives us
reasons to believe that the components are ellipsoidal. It seems
interesting to study how non-spherical Cepheids pulsate.
TYC~1031~01262~1 gives a rare opportunity to learn it.

\section*{Acknowledgments}

We would like to thank N.N. Samus for useful discussion of the
results and V.P. Goranskij for software we used in our work. This
study made use of the ASAS and NSVS projects data. Two of the
authors (S. Antipin and K. Sokolovsky) are grateful to the Russian
Foundation of Basic Research (grants No. 05-02-16688 and No.
05-02-16289) for partial support of this study.

\label{lastpage}

\end{document}